\pdfoutput=1
\documentclass{article}
\usepackage{graphicx}
\usepackage{harvard, authblk, indentfirst, upgreek, hyperref}
\usepackage[margin = 3cm]{geometry}

\begin{document}
\title{%
	\emph{decryst}: an efficient software suite for
	structure determination from powder diffraction%
}
\author{%
	Yu Liu \texttt{<caspervector@gmail.com>}\\
	National Laboratory for Superconductivity,
	Institute of Physics, Chinese Academy of Sciences,
	Beijing 100190, People's Republic of China;
	University of Chinese Academy of Sciences,
	Beijing 100149, People's Republic of China%
}
\date{}
\maketitle

\section*{Synopsis}

Keywords: incremental computation; parallel and distributed computing;
powder diffraction; structure determination; computer program.

\emph{decryst} is an open source software suite for structure determination
from powder diffraction using the direct space method, which can apply
anti-bump constraints automatically in global optimisation using the efficient
algorithm by \citeasnoun{liu2017}.  \emph{decryst} can offer high performance
due to the application of incremental computation, and is designed with parallel
and distributed computing in mind.

\section*{Abstract}

Presented here is \emph{decryst}, a software suite for structure determination
from powder diffraction, which uses the direct space method, and is able to
apply anti-bump constraints automatically and efficiently during the procedure
of global optimisation using the crystallographic collision detection algorithm
by \citeasnoun{liu2017}.  \emph{decryst} employs incremental computation in its
global optimisation cycles, which results in dramatic performance enhancement;
it is also designed with parallel and distributed computing in mind, allowing
for even better performance by simultaneous use of multiple processors.
\emph{decryst} is free and open source software, and can be
obtained at \url{https://gitlab.com/CasperVector/decryst/};
it strives to be simple yet flexible, in the hope that the underlying
techniques could be adopted in more crystallographic applications.

\section{Introduction}

In structure determination from powder diffraction (SDPD) \cite{david2002},
the reciprocal space methods, \emph{eg.}\ direct methods (not to be
confused with the direct space method), Patterson synthesis and charge
flipping \cite{baerl2006}, work by extracting the amplitudes of structure
factors for individual reflections and then phasing the reflections;
they usually require high resolution diffraction data, which can be
difficult or impossible to obtain under certain circumstances.
In contrast, the direct space method \cite{cerny2007} works by using global
optimisation algorithms to find a coordinate combination for independent
atoms in the unit cell that minimises an objective function, with the most
important part of the function being the divergence between the originally
observed diffraction pattern and the pattern computed from the coordinates,
usually measured by the Bragg $R$ factor; it is better at processing low
resolution data and solving molecular crystals.

However, as noted by \citeasnoun{liu2017}, even for structures as simple as
PbSO$_4$, there can exist several obviously unreasonable crystallographic
models with $R$ factors comparable to (or even smaller than) that of
the correct model, and most of them have some bumping atom pairs.
In the direct space method, it is desirable to automatically eliminate
these unreasonable models during the procedure of global optimisation,
which requires real-time crystallographic collision detection.
\citeasnoun{liu2017} presented a generic algorithm for this,
and proposed an evaluation function for atom bumping;
based on this research, we developed \emph{decryst},
a SDPD software suite, which is the subject of this article.

\begin{figure}[htbp]\centering
\includegraphics[width = 0.6\textwidth]{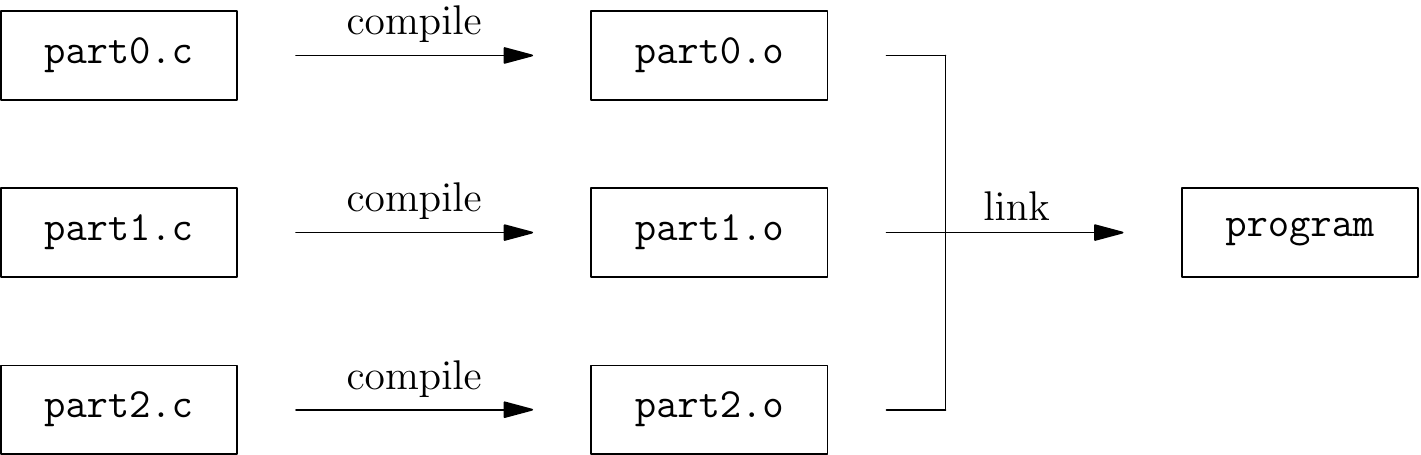}
\caption{%
	How \emph{make} builds an example program: after \texttt{program} is built,
	to rebuild it when \texttt{part0.c} somehow changes, \emph{make} would only
	recompile \texttt{part0.c} and then relink all \texttt{.o} files%
}\label{fig:make-build}
\end{figure}

Inspired by the \emph{make} utility \cite{feldman1979}, which identifies the
pieces of a project that have changed, and then executes commands to rebuild
these pieces (Figure \ref{fig:make-build}), we applied incremental computation
in \emph{decryst}, resulting in a dramatic speedup of the global optimisation
procedure.  But the speedup from incremental computation is still limited when
there are too many independent coordinates to be determined, because the size
of the search space grows exponentially with regard to the degree of freedom
(DOF).  Since computers with multiple processors are the norm today, we designed
\emph{decryst} with parallel and distributed computing in mind, to further
enhance its performance by employing multiple processors simultaneously.

\section{Underlying techniques in \emph{decryst}}

\subsection{\emph{decryst}: an overview}\label{ssec:oview}

\emph{decryst} is a software suite that attempts to determine the structure of
crystals from their indexed powder diffraction data using the typical direct
space method summarised by \citeasnoun{cerny2007}.  It also employs the
equivalent position combination (EPC) method \cite{deng2009}, so that it can
transform the original $3n$-dimensional global optimisation problem into
multiple smaller and mutually independent optimisation problems: \emph{eg.}\ for
PbSO$_4$ ($Pnma$, $Z = 4$), \emph{decryst} transforms the 72-dimensional
optimisation problem into 35 small optimisation problems, each corresponding to
a single EPC with the DOF ranging from 6 to 12.

Given this background, it is expected that \emph{decryst} would be functionally
similar to \emph{EPCryst} \cite{deng2011}; however, \emph{decryst} differs from
\emph{EPCryst}, most importantly in that it can automatically apply anti-bump
constraints using the efficient crystallographic collision detection algorithm
by \citeasnoun{liu2017}.  Additionally, the computational performance of
\emph{decryst} is greatly improved over \emph{EPCryst} through, to our
knowledge, the first generic application of incremental computation in
the global optimisation procedure (\emph{cf.}\ Subsection \ref{ssec:incr}).
Moreover, \emph{decryst} is designed with parallel and distributed computing
in mind, allowing for further enhancement of the performance
(\emph{cf.}\ Subsection \ref{ssec:para}).

We would like to note that, even without employing incremental computation
or parallelism, \emph{decryst} is already vastly faster than \emph{EPCryst}:
\emph{eg.}\ on our computer, global optimisation for the correct EPC of PbSO$_4$
requires several seconds with \emph{decryst}, and more than 1000 seconds with
\emph{EPCryst}.  This is achieved through (1) frugal implementation of
algorithms, (2) application of cache-friendly algorithms and data structures
\cite{bryant2011}, (3) use of an effective adaptive simulated annealing
schedule \cite{lam1988}, and (4) fast computation of structure factors using
the formulae (where $\vec{q}$ is the scattering vector):
\[
	F(\vec{q}) = \sum_i f_i(\vec{q}) (
		\mathrm{e} ^ {2\uppi\mathrm{i}\vec{q} \cdot \vec{x}_i} +
		\mathrm{e} ^ {2\uppi\mathrm{i}\vec{q} \cdot (-\vec{x}_i)}
	) = 2\sum_i f_i(\vec{q}) \mathrm{cos}(2\uppi\vec{q} \cdot \vec{x}_i)
\]
(for centrosymmetric cells, where $i$ is the index of a
centrosymmetric atom pair), and
\[
	F(\vec{q}) = \sum_{ij} f_i(\vec{q})
		\mathrm{e} ^ {2\uppi\mathrm{i}\vec{q} \cdot (\vec{x}_i + \vec{r}_j)}
	= \sum_i f_i(\vec{q}) \mathrm{e} ^ {2\uppi\mathrm{i}\vec{q} \cdot \vec{x}_i}
		\sum_j \mathrm{e} ^ {2\uppi\mathrm{i}\vec{q} \cdot \vec{r}_j}
\]
(for centred cells, where $\vec{r}_j$ is the coordinates of an origin).

\subsection{Incremental computation of the objective function}\label{ssec:incr}

\emph{decryst} uses a linear combination of the Bragg factor
\[
	R = \sum_{hkl} |I_{\mathrm{obs}, hkl} - I_{\mathrm{calc}, hkl}|
		\Big/ \sum_{hkl} I_{\mathrm{obs}, hkl}
\]
and an evaluation function $B$ for atom bumping as the objective function
\[ E = \mu B + (1 - \mu) R / 2, \]
where $\mu \in [0, 1]$ is the combination factor \cite{liu2017}.
So obviously, in order to speed up the global optimisation procedure, we need
to speed up the computation of $R$ and $B$; and since $R$ most importantly
depends on structure factors, to speed up the computation of $R$, we need to
speed up the computation of structure factors.

Often, the difference between crystallographic models in two adjacent global
optimisation cycles is simply the displacement of one independent atom,
so the structure factors in the two cycles are,
by additivity of the Fourier transform, subject to the relation
\[
	F'(\vec{q}) - F(\vec{q}) = \sum_j f_j(\vec{q}) (
		\mathrm{e} ^ {2\uppi\mathrm{i}\vec{q} \cdot \vec{x}'_j} -
		\mathrm{e} ^ {2\uppi\mathrm{i}\vec{q} \cdot \vec{x}_j}
	),
\]
where $j$ is the index of a moved atom.  Using this relation,
we are able to incrementally compute structure factors using
previously computed structure factors: we only need to recompute the
$\mathrm{exp}(2\uppi\mathrm{i}\vec{q} \cdot \vec{x}_j)$ term for the moved
atoms instead of for all atoms in the unit cell; therefore if there are $m$
independent atoms in the unit cell, and the atoms are moved in turn between
the optimisation cycles, recomputation for all atoms would be distributed into
$m$ cycles (instead of only one cycle), resulting in $m$ times performance
in computation of the structure factor compared with the na\"ive approach.

Similarly, for atom bumping evaluation functions based on a two-body potential
\[ C = \sum_{\{k_0, k_1\}} c(k_0, k_1), \]
where $c$ is a given two-body function and $k$ is the index of an atom in
the unit cell, it is trivial that $c(k_0, k_1)$ is only changed for pairs
$\{k_0, k_1\}$ involving a moved atom, so $C$ can also be computed
incrementally: we only need to recheck the collision between these pairs.
Remembering that by exploiting the equivalent position symmetry \cite{liu2017},
we only need to check pairs involving atoms in the asymmetric unit,
so the total number of re-tests is $(n - 1) + (m - 1) (n' - 1)$
(where $n$, $m$ and $n'$ are the total number of atoms in the unit cell, the
total number of atoms in the asymmetric unit and the number of equivalent
atoms of the moved atom, respectively), which would be approximately $O(n)$
for the average case.  Therefore, by employing incremental computation,
we no longer need a dedicated broad phase, like sweep and prune,
in order to avoid $O(n^2)$ time complexity in collision detection,
since using the equivalent position symmetry is already sufficient.
The evaluation function $B$ in \emph{decryst} is based on
a two-body potential, so this technique is applicable.

We would like to note that when the number of optimisation cycles is large,
the accumulation of summation errors from floating point arithmetic can become
a serious problem.  Incremental floating point summation algorithms like the
one by \citeasnoun{kahan1965} are inapplicable, since the mean value of the
numbers being summed is asymptotically zero \cite{higham1993}.  For now,
\emph{decryst} works around this by internally using fixed point arithmetic,
which is free from summation errors, for incremental summations.

\subsection{Performance evaluation of incremental computation in \emph{decryst}}

Apart from those data available at the project homepage
\url{https://gitlab.com/CasperVector/decryst/}, additional test data
for this article can be obtained from the supplementary materials.
All time consumptions reported in this subsection
were measured on an Intel i7-3720QM CPU.

We obtained several crystal structures of varying complexities from
the American Mineralogist Crystal Structure Database \cite{downs2003}.
For each structure, we computed the objective function $E$, either
with or without using incremental computation (corresponding to
$t_{\cdots,\mathrm{incr}}$ and $t_{\cdots,\mathrm{orig}}$ in the table,
respectively), for 100 groups of crystallographic models, each group consisting
of 100 models with random coordinates of the independent atoms.  We collected
the averages and standard deviations of time consumed in computation of either
the entire objective function ($t_{E,\cdots}$) or only the Bragg $R$ factor
($t_{R,\cdots}$) for each group divided by the number of models
in each group, as shown in Table \ref{tbl:incr-eval}.

\begin{table}[htbp]\centering
\caption{%
	Results for performance evaluation of incremental computation
	in \emph{decryst} (all time consumptions are in $\upmu$s):
	``ID'' is the AMCSD database code for the test structure, $n$ is the
	total number of atoms in the unit cell, $m$ is the number of independent
	atoms, $N_\mathrm{refl}$ is the number of recorded reflections;
	$t_{E,\cdots}$ and $t_{R,\cdots}$ are time consumed in
	computation of $E$ and $R$ for each model, respectively;
	$t_{\cdots,\mathrm{incr}}$ and $t_{\cdots,\mathrm{orig}}$ are
	obtained with and without using incremental computation, respectively
}\label{tbl:incr-eval}\vspace*{1em}
\begin{tabular}{cccccccc}\hline
ID &	$n$ &	$m$ &	$N_\mathrm{refl}$ &	$t_{R,\mathrm{incr}}$ &
$t_{R,\mathrm{orig}}$ &	$t_{E,\mathrm{incr}}$ &	$t_{E,\mathrm{orig}}$ \\\hline
0005558 &	24& 5& 94 &	8.9(3) &	43.9(5) &	12.5(7) &	44.5(7) \\
0009272 &	64& 8& 80 &	39.7(9) &	241(3) &	50.6(6) &	242(3) \\
0009563 &	90& 10& 76 &	15.0(2) &	168(3) &	30.0(7) &	174(2) \\
0002630 &	126& 14& 73 &	11.3(3) &	244(3) &	30.7(3) &	252(2) \\
0000427 &	152& 20& 374 &	27.4(5) &	639(7) &	53.6(7) &	663(8) \\
0000447 &	160& 4& 32 &	27.0(9) &	87(2) &	59(1) &	102(2) \\\hline
\end{tabular}
\end{table}

From the table we conclude that for many crystal structures, computation
of the objective function, including both the Bragg $R$ factor and the
atom bumping evaluation function, can be dramatically (sometimes by an order
of magnitude or more) sped up by incremental computation.  We note that
the time consumptions for ``0009272'' and $t_{\cdots,\mathrm{orig}}$ for
``0000447'' might appear anomalous, but actually the former can be explained
by the fact that the structure does not have a centred cell and that none of
the used Wyckoff positions are centrosymmetric (\emph{cf.}\ Subsection
\ref{ssec:oview}), while the latter can be explained by the small $m$
in comparison with $n$ \cite{liu2017} and the much smaller $N_\mathrm{refl}$
in comparison with those of all other test structures.

\subsection{Parallelism with \emph{decryst}}\label{ssec:para}

\begin{figure}[htbp]\centering
\includegraphics[width = 0.9\textwidth]{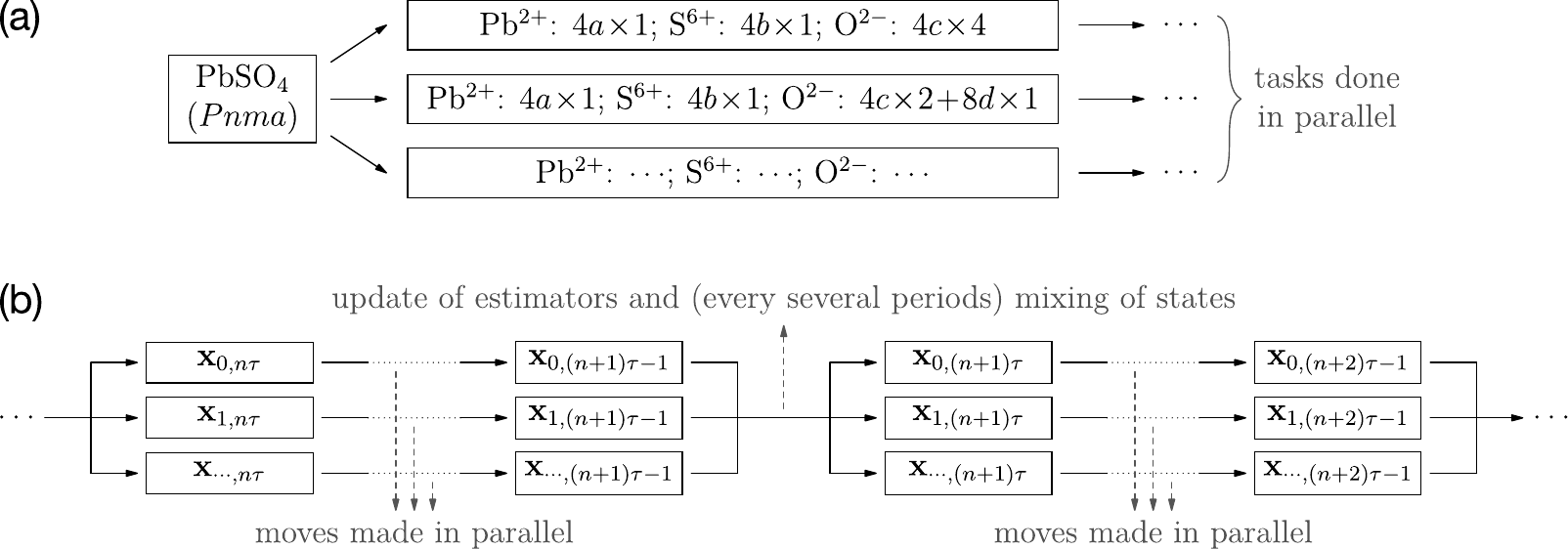}
\caption{%
	Two types of parallelism in \emph{decryst}:
	(a) parallel processing of tasks on different EPCs;
	(b) parallelisation of the simulated annealing algorithm%
}\label{fig:decr-para}
\end{figure}

Just like \emph{EPCryst}, \emph{decryst} filters out obviously unreasonable
EPCs using statistical analyses, and then performs global optimisation on the
remaining ones; and since tasks on different EPCs are mutually independent,
\emph{decryst} performs them in parallel (Figure \ref{fig:decr-para}(a)).
As a result from parallel processing of tasks on different EPCs,
\emph{decryst}'s ability for structure determination is greatly improved
over \emph{EPCryst}.  As \citeasnoun{deng2011} noted, ``\emph{EPCryst} works
particularly well when the number of generated TSMs is small (say less than
100) and the number of parameters is not greater than ten'' (a ``TSM'' is
a trial structure model corresponding to a single EPC); in contrast, by
distributing the tasks among multiple processors, and applying previously
discussed performance optimisation techniques, \emph{decryst} can handle
thousands of EPCs, each with a DOF more than 20.  Noticing the large number
of possible EPCs and the dramatic reduction in the DOF resulted from the EPC
method in most useful cases, we expect the parallelisation of EPC tasks to
offer huge opportunities.

For EPCs with large DOFs, \emph{decryst} can also perform parallel global
optimisation: it uses the parallel simulated annealing algorithm by
\citeasnoun{chu1999}, which is based on the adaptive annealing schedule by
\citeasnoun{lam1988}.  The Lam schedule uses continuous cooling, with the
cooling rate dynamically computed from several statistical estimators
gathered periodically (every $\tau$ cycles); for best performance, the average
move size is also dynamically controlled, to keep the acceptance ratio of the
moves around 0.44.  \citeasnoun{chu1999} noted that the underlying physical
metaphor of simulated annealing is the sampling of Boltzmann distributions,
and that multiple sampling processes can be correlated by periodic mixing
of the latest states from all processes (Figure \ref{fig:decr-para}(b)),
with the probability for choosing a state under temperature $T$ being
\[ p_j = \mathrm{exp}(-E_j / T) \Big/ \sum_j \mathrm{exp}(-E_j / T), \]
where $E_j$ is the value of the objective function on process $j$.
Using this technique, the sampling procedure can be distributed among
multiple processes, hence enabling parallelism in simulated annealing.

For our application, we made some minor adjustments to the original algorithms:
\begin{itemize}
\item Due to the periodic boundary of the unit cell, we model the search
	domain of the coordinate combination as a hyperrectangle with a periodic
	boundary instead of an infinite search domain as in the original algorithms.
	Accordingly, the move size is generated from a random-signed wrapped
	exponential distribution instead of a random-signed exponential
	distribution.
\item Since the coordinate combination is randomly initialised, initial cycles
	at infinite temperature used to erase the initial state are no longer
	needed.  Similarly, because we have a dedicated statistical analysis stage
	for EPCs, initial cycles used to gather statistics for initialisation of the
	statistical estimators are no longer necessary.
\end{itemize}

\section{Applications of \emph{decryst}}\label{sec:appl}

\subsection{\emph{decryst}: the programs}

\emph{decryst} is free and open source software that
runs on Unix-like operating systems, and can be obtained
at \url{https://gitlab.com/CasperVector/decryst/}.
Realising that we cannot afford to ``reinvent the wheel'' with our limited
resource, we implemented \emph{decryst} in a simple yet flexible way.
We explicitly wish the techniques in \emph{decryst} to be examined by
fellow crystallographers, and eventually be adopted in more applications;
we believe this will hopefully result in significantly better performance
of crystallographic software in general.

\begin{figure}[htbp]\centering
\includegraphics[height = 0.8\textheight]{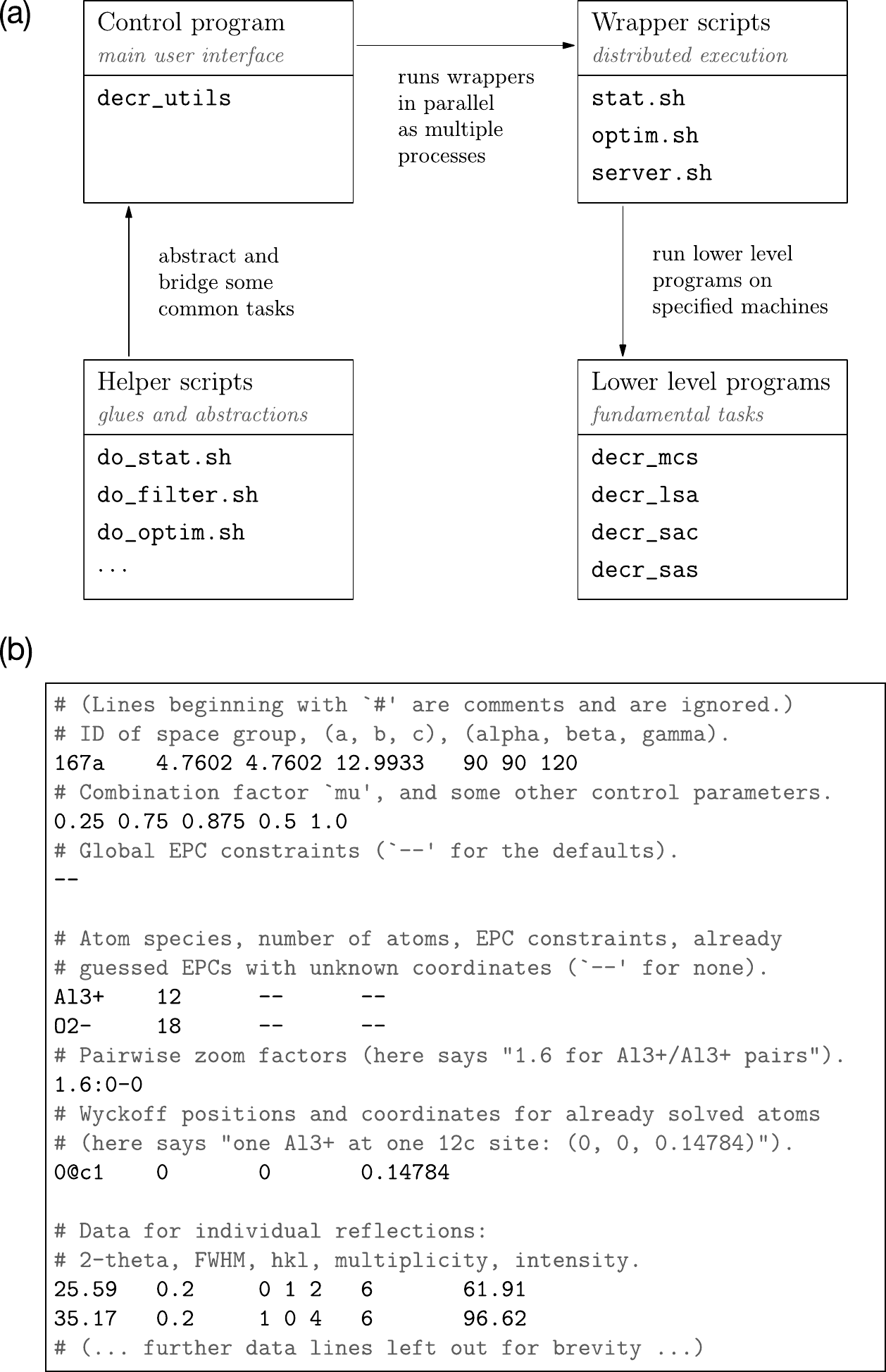}
\caption{%
	(a) Organisation of individual programs in \emph{decryst};
	(b) an example input file for \emph{decryst}'s control program
	(for Al$_2$O$_3$, with AMCSD database code ``0009325'')%
}\label{fig:decr-desc}
\end{figure}

\emph{decryst} is composed of multiple parts (Figure \ref{fig:decr-desc}(a);
details on invocation of individual programs and their input formats
have been provided at the project homepage, and can be freely accessed):
\begin{itemize}
\item Several single-purpose lower level programs, each of which implements
	one basic functionality (statistical analyses, global optimisation);
	they are written in C to achieve high performance.  These programs and
	the control program introduced below comprise the core of \emph{decryst}.
\item One higher level multi-purpose program that controls the lower level
	programs; this program is written in Python for maximal flexibility.
	Some functionalities (and in particular, enumeration of EPCs),
	which are fairly complex but not performance-critical,
	are also implemented in this program.
\item Wrapper scripts that invoke a lower level program on a specified machine
	according to some control parameters.  When the control program performs
	a task, it actually runs wrapper scripts in parallel, resulting in specified
	lower level programs to be executed on specified machines in parallel, hence
	enabling parallel and distributed computing.
\item Helper scripts that bridge different tasks and provide higher
	abstractions, \emph{eg.}\ one that filters the list of EPCs for use in
	global optimisation according to the results of statistical analyses using a
	figure of merit like the one by \citeasnoun{deng2011}.  Helpers and wrappers
	are written as Unix shell scripts.
\end{itemize}

\emph{decryst} uses textual interfaces (\emph{eg.}\ the input format for the
control program as shown in Figure \ref{fig:decr-desc}(b)) that are friendly
to standard Unix text processing utilities.  This design allows \emph{decryst}
to perform highly automated tasks exactly at the will of the user; as an example
for this, we solved the ``0009563'' structure in Table \ref{tbl:incr-eval}.
All time consumptions reported in this section were measured on an Intel
i7-3720QM CPU; the procedure has been provided at the project homepage,
and detailed results can be obtained from the supplementary materials.

\subsection{``0009563'': solving the heavy atoms}

From the indexed XRD data (\emph{decryst} merges the intensities for reflections
that overlap in $2\theta$, so the data were treated as regular powder
diffraction data) for ``0009563'' obtained from AMCSD, we are able to tell that
the formula for its $P\bar{3}c1$ unit cell is Rb$_{12}$Ti$_6$Ge$_{18}$O$_{54}$.
Since all metal atoms in this structure are much heavier than O$^{2-}$,
it is suitable to solve the structure using the heavy atom method.
After commenting out lines in the input file involving the O$^{2-}$ atoms
(\emph{cf.}\ Figure \ref{fig:decr-desc}(b)), we instructed \emph{decryst}
to enumerate all possible EPCs for the heavy atoms according to
the input file, resulting in 1451 EPCs in total.

Before attempting to actually solve the heavy atoms, it is very useful to
filter out the unreasonable EPCs using certain criteria.  In early works,
\emph{eg.}\ those by \citeasnoun{lu1965} and \citeasnoun{reddy1965}, for
simple crystals, crystallographers were able to identify EPCs that are bound to
produce crystallographic models with atom bumping regardless of the independent
atoms' coordinates by skillful analyses of the geometric properties of the EPCs.
Using the atom bumping evaluation function $B$ (\emph{cf.}\ Subsection
\ref{ssec:incr}), we can perform the same task programmatically in
\emph{decryst}: by setting the combination factor $\mu$ to 0.99 (we deliberately
avoided using 1 exactly, so that the optimisation algorithm would not terminate
too early with EPCs that produced models with $B = 1$ with a big probability)
before running global optimisation on the EPCs, we were able to find those EPCs
that were incapable of producing models free from bumping at all.

After examining the results from the previous global optimisation step,
we dropped all EPCs with final $E > 0.12$, leaving 183 (13\%) EPCs as
candidates for the next step.  We then restored the original $\mu$ for
these EPCs, and again performed global optimisation on them; since these EPCs
had DOFs ranging from 5 to 8, and only involved heavy atoms, the results were
quite reproducible, so we only ran global optimisation once for each EPC.
After examining the results, we discarded the EPCs with final $E > 0.2$,
leaving 26 EPCs for the heavy atoms; noticing that the $B$ values for
these EPCs clearly fell into two different groups, one with the greatest
$B = 0.0027$ and the other with $B$ ranging from 0.056 to 0.112,
we dropped all EPCs with $B > 0.02$, leaving 20 (11\%) of them.

By enabling parallelism in \emph{decryst} (achieved by specifying appropriate
control parameters in a ``hosts'' file), we were able to make full use of all
8 available CPU threads.  With parallelism enabled, the procedures (excluding
human intervention) in this subsection took about 2.5 minutes to complete.
As a matter of fact, the heavy atoms were solved correctly in this step: the
correct EPC (Rb$^+$: $12g\times1$; Ti$^{4+}$: $2b\times1+4d\times1$; Ge$^{4+}$:
$6f\times1+12g\times1$) had the smallest $E = 0.058$, and was solved correctly;
visual inspection of solutions for the 6 EPCs rejected due to large $B$ values
confirmed that they all had obviously bumping atom pairs.

\subsection{``0009563'': solving the O$^{2-}$ atoms}\label{ssec:oxygen}

\emph{decryst} can automatically merge solutions with the input file to
produce new input files for solved EPCs with all comments preserved, as well as
CIF files for the (partially) solved structures.  After merging the solutions
for the candidate EPCs from the previous step, we uncommented the O$^{2-}$ lines
in all these input files, and enumerated full EPCs based on the previously
chosen EPCs for the heavy atoms, resulting in 4455 EPCs in total.  Then again we
attempted to identify the unreasonable ones among these EPCs using the function
$B$ by temporarily setting $\mu = 0.99$: in order to save time, we filtered out
all EPCs with smallest $E > 0.99$ after the statistical analyses step,
resulting in only 2186 (49\%) EPCs remaining; after the global
optimisation step, we dropped the EPCs with final $E > 0.05$,
leaving 110 (5\%) candidates for the next step.

Considering that the DOFs for these EPCs were mainly 10 to 13, and that O$^{2-}$
has very limited contribution to the diffraction pattern of ``0009563'', the
results from global optimisation on these EPCs were expected to vary from run to
run, so we needed to perform global optimisation several times on the EPCs and
pick the plausible solutions.  In order to reduce the number of candidates for
visual inspection, we gradually reduced the number of results after the runs:
after the first run, we dropped all EPCs with final $B > 0.02$, leaving 50 EPCs;
after the second run, we chose the 20 EPCs with smallest $E$ values; after the
third run, we chose the top 10 EPCs.  It was retroactively confirmed that the
final candidates usually had the smallest $E$ values in these runs.

\begin{figure}[htbp]\centering
\includegraphics[height = 0.84\textheight]{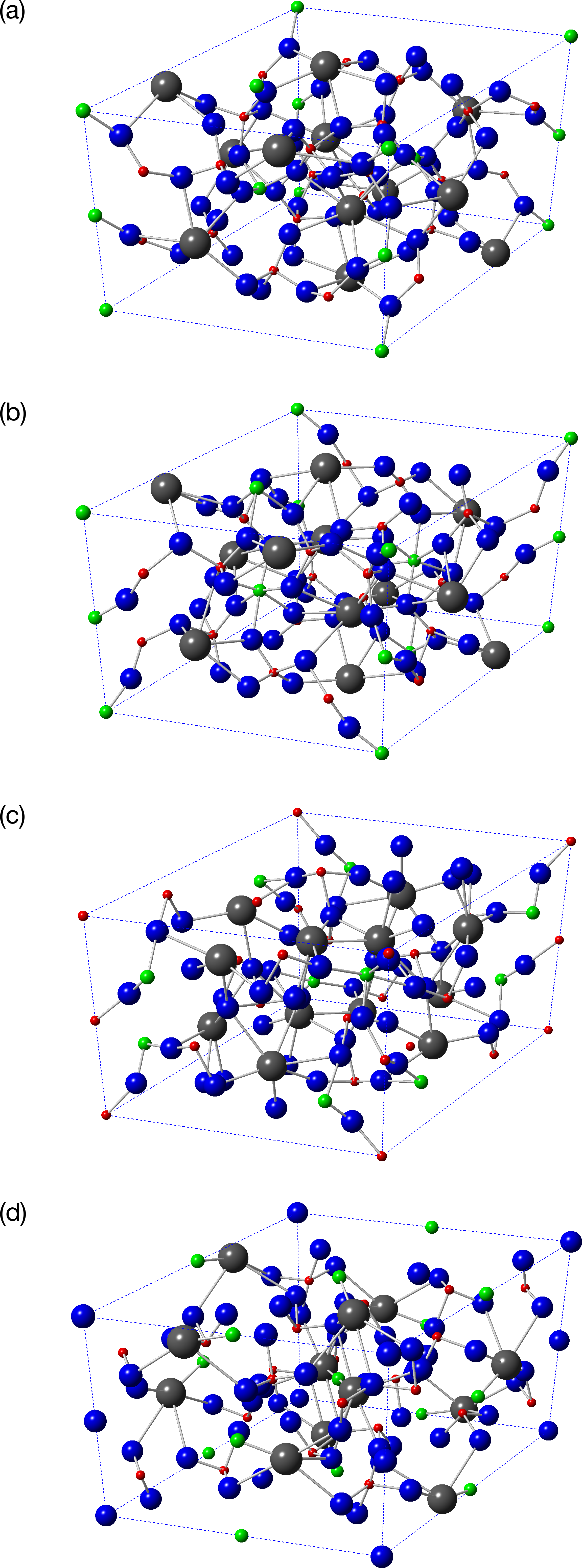}
\caption{%
	Some solutions for ``0009563'': (a) the one closest to the correct solution,
	from EPC [Rb$^+$: $12g\times1$; Ti$^{4+}$: $2b\times1+4d\times1$; Ge$^{4+}$:
	$6f\times1+12g\times1$; O$^{2-}$: $6f\times1+12g\times4$]; (b) the other
	plausible one, from the same EPC as (a), but with painful coordination
	polyhedra for the Ge$^{4+}$ atoms; (c) one with each Ge$^{4+}$ atom
	on a cell edge coordinated by 6 O$^{2-}$ atoms and each of the rest
	coordinated by 3 O$^{2-}$ atoms, from EPC [Rb$^+$: $12g\times1$; Ti$^{4+}$:
	$6f\times1$; Ge$^{4+}$: $2b\times1+4d\times1+12g\times1$; O$^{2-}$:
	$6f\times1+12g\times4$]; (d) one with each Ti$^{4+}$ atom coordinated
	by 2 Rb$^+$ atoms in the presence of lone O$^{2-}$ atoms, from EPC
	[Rb$^+$: $12g\times1$; Ti$^{4+}$: $6e\times1$;
	Ge$^{4+}$: $6f\times1+12g\times1$; O$^{2-}$:
	$2b\times1+4d\times1+6f\times2+12g\times3$]%
}\label{fig:rgto-sol}
\end{figure}

For each of the 10 remaining EPCs, we performed global optimisation 10 times;
among the 100 solutions, we discarded those with $B > 0.02$, leaving 90 of them,
and then visually inspected the remaining solutions.  Crude inspection showed
that the correct EPC (with O$^{2-}$ at $6f\times1+12g\times4$) was a clear
winner, since all solutions from other EPCs had ill-defined bonding relations:
patently uneven coordination numbers for the same kind of atoms (Figure
\ref{fig:rgto-sol}(c)), metal atoms occupying sites with too many metallic
neighbours in the presence of lone O$^{2-}$ atoms in the unit cell (Figure
\ref{fig:rgto-sol}(d)), \emph{etc}.  Noticing that this EPC also produced
the 10 solutions with smallest $E$ values among all 100 solutions,
it was without doubt the only plausible EPC for ``0009563''.

Among the 10 solutions from this EPC, we found two distinct plausible ones:
one of them (Figure \ref{fig:rgto-sol}(a)) was very close to the originally
reported structure, and only needed some refinement; the other (Figure
\ref{fig:rgto-sol}(b)) did not appear immediately incorrect, but Ge$^{4+}$
atoms in the structure had quite painful coordination polyhedra.
With parallelism enabled, the procedures (excluding human intervention)
in this subsection took 25 minutes in total.

\section{Discussion and conclusion}

\subsection{Discussion}

Incremental computation of objective functions, including the more elaborate
ones (\emph{eg.}\ those involving coordination numbers and valences), should
be readily applicable to programs that mainly use atomic coordinates as the
structural parameters for the optimisation algorithm, \emph{eg.}\ \emph{FOX}
\cite{favre2002} and \emph{FraGen} \cite{li2012}; nevertheless, this technique
is perhaps not trivially usable in direct space method programs that mainly
specialise in molecular crystals, \emph{eg.}\ \emph{EXPO} \cite{altomare2013},
because of the heavy use of bond lengths, bond angles, \emph{etc.}\ in the
independent variables, where even a single change in one parameter might
affect all atomic coordinates.

Even though most crystallographic software do not employ the EPC method,
parallelisable tasks are widespread: \emph{eg.}\ for SDPD of complicated
structures, it is often necessary to run the optimisation algorithm several
times before picking a best result (\emph{cf.}\ Subsection \ref{ssec:oxygen}),
and these runs are trivially parallelisable, similar to the situation in
Subsection \ref{ssec:para}.  Parallelisation of optimisation algorithms can
also be useful: \emph{eg.}\ by parallelising the algorithms used in structure
refinement, we would be able to dramatically reduce the time spent by
the user on waiting for the results; we note that the recent work by
\citeasnoun{eremenko2017} is already in this direction.

As demonstrated in Section \ref{sec:appl}, apart from \emph{a posteriori}
structure validation and real-time elimination of crystallographic models with
atom bumping, the atom bumping evaluation function can also be used in automatic
\emph{a priori} filtering of EPCs that cannot produce reasonable models at all.
Filtering using other evaluation functions, \emph{eg.}\ those that evaluate
the goodness of bonding relations in crystallographic models, should also be
usable in principle, and will surely be of interest: \emph{eg.}\ with such
advanced filtering, the manual steps in Subsection \ref{ssec:oxygen} could be
greatly simplified.

As mentioned in Section \ref{sec:appl}, even with \emph{a priori} filtering
of EPCs that would produce unreasonable crystallographic models, and with the
objective function employing anti-bump constraints, we might still get solutions
with atom bumping.  We think this is because certain EPCs just cannot produce
models with small Bragg $R$ factors without introducing atom bumping, or can but
only with a very small probability.  For this reason, we still need to perform
\emph{a posteriori} filtering of solutions with atom bumping.

\subsection{Conclusion}

We presented \emph{decryst}, an open source software suite for crystal
structure determination from powder diffraction data.  \emph{decryst} can
offer high performance due to the application of incremental computation,
and the performance can be further enhanced by using its features for parallel
and distributed computing.  \emph{decryst} is simple yet flexible, and we
explicitly wish the underlying techniques to be adopted in more
crystallographic applications.

\section*{Acknowledgements}

The author is deeply indebted to all users that participated in the testing
of \emph{decryst}, whose suggestions were invaluable to the documentation of
the software.  The author would also like to thank Cheng Dong for bringing
up the issue of crystallographic collision detection, thank the Bilbao
Crystallographic Server \cite{aroyo2006} for kindly providing
crystallographic data in the spirit of open access, and thank the anonymous
referees for their enlightening comments.  This article is dedicated to
Pieter Hintjens (1962.12.3 -- 2016.10.4), a main author of \emph{ZeroMQ},
the message passing library used in the implementation of parallel simulated
annealing in \emph{decryst}.  This project was supported by the National
Natural Science Foundation of China under grant No.\ 21271183.

\bibliographystyle{agsm}
\bibliography{art2}
\end{document}